\documentclass[a4paper,twocolumn,english,,prb,showpacs]{revtex4}
\usepackage[T1]{fontenc}
\usepackage[latin9]{inputenc}
\usepackage{float}
\usepackage{amsmath}
\usepackage{graphicx}

\makeatletter

\providecommand{\tabularnewline}{\\}

\@ifundefined{textcolor}{}
{%
 \definecolor{BLACK}{gray}{0}
 \definecolor{WHITE}{gray}{1}
 \definecolor{RED}{rgb}{1,0,0}
 \definecolor{GREEN}{rgb}{0,1,0}
 \definecolor{BLUE}{rgb}{0,0,1}
 \definecolor{CYAN}{cmyk}{1,0,0,0}
 \definecolor{MAGENTA}{cmyk}{0,1,0,0}
 \definecolor{YELLOW}{cmyk}{0,0,1,0}
 }

\makeatletter



\makeatother

\makeatother

\usepackage{babel}

\makeatother

\usepackage{babel}

\begin{document}

\title{An \emph{ab initio} study of $3s$ core-level x-ray photoemission
spectra in transition metals}

\author{Manabu Takahashi$^{1}$ and Jun-ichi Igarashi$^{2}$ }

\affiliation{$^{1}$Faculty of Engineering, Gunma University, Kiryu, Gunma 376-8515,
Japan\\
$^{2}$Faculty of Science, Ibaraki University, Mito, Ibaraki 310-8512,
Japan}
\begin{abstract}
We calculate the $3s$- and $4s$-core-level x-ray photoemission spectroscopy
(XPS) spectra in the ferromagnetic and nonmagnetic transition metals
by developing an \emph{ab initio} method. We obtain the spectra exhibiting
the characteristic shapes as a function of binding energy in good
agreement with experimental observations. The spectral shapes are
strikingly different between the majority spin channel and the minority
spin channel for ferromagnetic metals Ni, Co, and Fe, that is, large
intensities appear in the higher binding energy side of the main peak
(satellite) in the majority spin channel. Such satellite or shoulder
intensities are also obtained for nonmagnetic metals V and Ru. These
behaviors are elucidated in terms of the change of the one-electron
states induced by the core-hole potential.  
\end{abstract}


\maketitle

\section{Introduction}

X-ray spectroscopy has been extensively used for studying electronic
properties in solids. Core-level spectroscopy is particularly useful
for investigating the electronic states through the dynamical response
to the photo-created core-hole. It is well known that the response
function in metallic systems exhibits the singular behavior near the
Fermi edge.\cite{Anderson1967,Mahan1967,Nozieres1969} In the case
of the core-level x-ray photoemission spectroscopy (XPS), the spectra
display the asymmetric shapes as a function of binding energy in the
vicinity of the threshold.\cite{Doniach1970} Apart from the edge
singularity, some structures have been observed in the high binding-energy
region in some ferromagnetic transition metals and their compounds.
A notable example is a satellite peak on the $2p$ XPS in ferromagnetic
metal Ni, which is located around the region $6$ eV higher from the
threshold. \cite{Hufner1975} Feldkamp and Davis\cite{Feldkamp1980}
analyzed these XPS spectra by evaluating the overlap between the excited
states and the ground state, using a numerical method on the linear-combination-atomic-orbital
model. They clarified the origin of satellite as a combined effect
of the core-hole screening and the interaction between electrons.

As regards the $3s$ XPS, many experiments have already been carried
out on ferromagnetic transition metals and their compounds.\cite{Fadley1970,Acker1988,Hillebrecht1990,Campen1994,Xu1995,See1995Fe,See1995Ni,Lademan1997,Shabanova2001,Kamakura2006}
Several materials show satellite or shoulder structures as a function
of binding energy, which are interpreted as a result of the $3s$
level splitting due to the exchange interaction between the $3s$
electrons and the valence electrons in the polarized $3d$ states.
In some cases, they have been related to the local magnetic moment.\cite{Bagus1994}
On the other hand, having intensively investigated Fe $3s$ XPS in
various iron compounds, Acker et al.\cite{Acker1988} revealed that
only poor correlation exists between the satellite structures and
the magnetic moments. They also found that the Fe $3s$ XPS spectra
show the satellite structure even in some Pauli paramagnetic compounds.
Furthermore, having investigated the Mn and Fe $3s$ XPS spectra in
the insulating compounds, Oh et al.\cite{Oh1992} concluded that the
splitting between and main and satellite peaks dose not reflect the
$3d$ moment when the effect of the charge-transfer becomes important.

We have developed an \emph{ab initio} method to calculate the XPS
spectra by extending the theory of Feldkamp and Davis.\cite{Feldkamp1980,Taka2008FeXPS}
Here we briefly summarize the procedure of calculation. First, we
carry out the band structure calculation within the local density
functional approximation (LDA) to obtain the one-electron states in
the ground state. Next, instead of considering the system with only
one core hole in crystal, we consider a system of super-cells with
one core-hole per cell. Strictly speaking, we should consider the
former system for the XPS event, but the latter system is expected
to work better as increasing the cell size. These systems correspond
to a kind of impurity problem, where the local charge neutrality has
to be satisfied according to the Friedel sum rule. \cite{Friedel1958}
We carry out the band structure calculation based on the LDA, in which
the exchange and Coulomb interactions between the core electrons and
the conduction electrons and between the conduction electrons are
taken into account through the exchange-correlation potential. The
charge variation due to screening the core hole in the final states
is also taken into account within the super-cell approximation. To
guarantee the charge neutrality, we add one extra conduction electron
in each super-cell, and seek the self-consistent solution. Then, with
the calculated one-electron states, we discretize the momentum space
into finite number of points, and construct final states by distributing
electrons on these one-electron states. The final state with the lowest
excited energy is given by piling the same number of electrons into
low energy one-electron levels at each $\mathbf{k}$-point as that
in the ground state. We prepare the other final states by creating
one electron-hole (e-h) pair, two e-h pairs, and so on. Finally, we
calculate the XPS spectra by evaluating the overlaps between thus
obtained final states and the ground state with the help of the one-electron
wave functions.

The purpose of this paper is to systematically clarify the relation
between the spectral shapes and the screening process by calculating
the spin resolved $3s$ XPS spectra in a series of ferromagnetic metals
Ni, Co, and Fe. The usefulness of our \emph{ab initio} method is demonstrated.
We have already reported the spectra in ferromagnetic Fe in Ref. \onlinecite{Taka2008FeXPS}.
In these metals, spectral shapes have characteristic dependence on
elements and spin channels; the spectra have satellite or shoulder
in the majority spin channel, while the spectra show single peak structures
in the minority spin channel. Here we define the majority (minority)
spin channel by the process that the $3s$-core electron is photo-excited
to the vacuum state with the same spin as the majority (minority)
spin in the conduction band states. As far as we know, such spectra
have been analyzed only by using a single band Hubbard model, and
has been related to the $3s$-$3d$ exchange interaction.\cite{Kakehashi1984}
However, the model is, we think, too simple to compare the calculated
results quantitatively with the experimental data and to draw definite
conclusion.

Applying the \emph{ab initio} method, we calculate the spectra in
good agreement with the experimental observations. The screening effects
are quite different between the spin channels due to the exchange
interaction between the $3d$ electrons and the core hole. The $3d$
states are modified by the core-hole potential at the core-hole site,
and sometimes quasi-bound states are created near the bottom of the
$3d$ band. The e-h pair excitations from such quasi-bound states
to the empty states correspond to the satellite or shoulder intensities.
We find that the presence of the quasi-bound states is not sufficient
and the $3d$ bands have to be partially occupied in the ground state,
in order that the satellite or shoulder structure appears. These considerations
well explain the characteristics of the XPS spectra.

Furthermore, to clearly show that the presence of satellite or shoulder
has no direct relation to magnetic states, we calculate the $3s$
XPS spectra in nonmagnetic metals V and Cu, and the $4s$ XPS spectra
in nonmagnetic metal Ru. We obtain shoulders in the high binding-energy
region in V and Ru. On the other hand, we have no such structure in
Cu, although the \emph{localized} bound states are clearly created
below the bottom of the $3d$ band. We could explain these behaviors
in the same way as in the ferromagnetic metals.

The present paper is organized as follows. In Sec. II, we formulate
the XPS spectra with the \emph{ab initio} method. In Sec. III, we
present the calculated XPS spectra and discuss the behavior. The last
section is devoted to the concluding remarks.

\section{Procedure of calculation}

\subsection{Formula for XPS spectra}

We consider the situation that a core electron is excited to a high
energy state with energy $\epsilon$ by absorbing an x-ray photon
with energy $\omega_{q}$ and that the interaction between the escaping
photo-electron and the other electrons could be neglected. The probability
of finding a photo-electron with energy $\epsilon$ and spin $\sigma$
could be proportional to \begin{align}
 & I_{\sigma}^{{\rm XPS}}(\omega_{q}-\epsilon)=\nonumber \\
 & \qquad2\pi|w|^{2}\sum_{f}|\langle f|s_{\sigma}|g\rangle|^{2}\delta(\omega_{q}+E_{g}-\epsilon-E_{f}),\label{eq:xps}\end{align}
 where $w$ represents the transition matrix element from the core
state localized at a particular site to the state of photo-electron,
and is assumed to be independent of energy $\epsilon$ and spin $\sigma$.
The $s_{\sigma}$ is the annihilation operator of a relevant core
electron, which is assumed to have only spin $\sigma$ as the internal
degrees of freedom. The kets $|g\rangle$ and $|f\rangle$ represent
the ground state with energy $E_{g}$ and the final state with energy
$E_{f}$, respectively. We define $|f\rangle$ by excluding the photo-electron.
In the following calculation, we replace the $\delta$-function by
the Lorentzian function with the full width of half maximum (FWHM)
$2\Gamma_{s}$ with $\Gamma_{s}=1.0$ eV in order to take account
of the life-time broadening of the core level.

\subsection{Construction of final and initial states}

\begin{figure}[H]
\hfill{}\includegraphics[clip,scale=0.3]{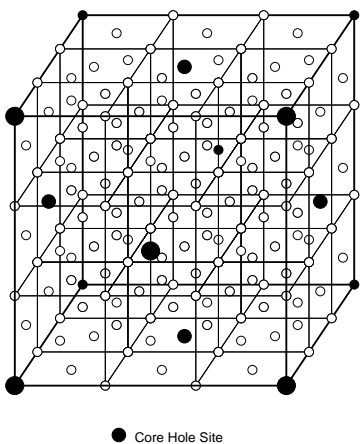}\hfill{}

\caption{Sketch of a super-cell containing core-holes in fcc structure. Core-hole
sites indicated by solid circles are assumed forming a $3\times3\times3$
fcc lattice. The super-cell for the bcc structure is shown in fig.
1 in Ref. \onlinecite{Taka2008FeXPS}. \label{fig:supercell}}

\end{figure}

In order to simulate the photo-excited states, we consider a periodic
array of super-cells with one core-hole per cell, and calculate the
one-electron states by means of the band structure calculation based
on the full potential linear augmented plane wave (FLAPW) method.
We use the $3\times3\times3$ bcc super-cell for Fe and V as shown
in Fig.~1 in Ref. \onlinecite{Taka2008FeXPS}, and the $3\times3\times3$
fcc super-cell for Co, Ni, Cu and Ru as shown in Fig. \ref{fig:supercell},
where the core-hole sites form a bcc lattice and an fcc lattice, respectively.
The larger the unit cell size is, the better results are expected
to come out. The $3s$- or $4s$-core states in transition metals
are treated as localized states within a muffin-tin sphere, so that
we could specify the core-hole site. To ensure the charge neutrality,
we assume $n_{e}+1$ band electrons per unit cell instead of $n_{e}$
band electrons, where $n_{e}$ is the number of band electrons per
cell in the ground state. One additional electron per unit cell would
not cause large errors in evaluating one-electron states in the limit
of large unit-cell size. The self-consistent potential is obtained
as the potential for the fully relaxed (screened) state. We write
the resulting one-electron state with energy eigenvalue $\epsilon_{\sigma n}(\mathbf{k})$
as\begin{equation}
\psi_{\sigma n\mathbf{k}}(\mathbf{r})=\frac{1}{\sqrt{N_{c}}}\sum_{j}\phi_{\sigma n\mathbf{k}}(\mathbf{r}-\mathbf{R}_{j})\exp(i\mathbf{k}\cdot\mathbf{R}_{j}),\end{equation}
 with $\phi_{\sigma n\mathbf{k}}(\mathbf{r})=u_{\sigma n\mathbf{k}}\left(\mathbf{r}\right)e^{i\mathbf{k}\cdot\mathbf{r}}$,
where $u_{\sigma n\mathbf{k}}(\mathbf{r})$ has the period of the
super-cell, and $j$ runs over $N_{c}$ super-cells. Needless to say,
wave vector ${\bf k}$'s have $N_{c}$ discrete values in the irreducible
Brillouin zone. We use these one-electron states as substitutes of
the states under a single core-hole. We distribute $n_{e}$ band electrons
per super-cell on these states to construct the excited states. In
addition, we carry out the band calculation in the absence of the
core-hole with assuming $n_{e}$ band electrons per super-cell. The
wave function and energy eigenvalue are denoted as $\psi_{\sigma n\mathbf{k}}^{(0)}(\mathbf{r})$
and $\epsilon_{\sigma n}^{(0)}(\mathbf{k})$, respectively. All the
lowest $N_{e}=N_{c}\times n_{e}$ levels are occupied in the ground
state.

The final states $\left|f\right\rangle $'s are constructed by using
the one-electron states calculated in the presence of the core-hole
in accordance with the following procedure. Defining $n_{\sigma}^{\left(0\right)}\left(\mathbf{k}\right)$
by the number of levels with spin $\sigma$ and wave vector $\mathbf{k}$
below the Fermi level in the ground state, we distribute $n_{\sigma}^{\left(0\right)}\left(\mathbf{k}\right)$
electrons with spin $\sigma$ and wave vector $\mathbf{k}$ in the
states given in the presence of core hole. The final state $\left|f_{0}\right\rangle $
containing no e-h pair is constructed by distributing electrons from
the lowest energy level up to the $n_{\sigma}^{\left(0\right)}\left(\mathbf{k}\right)$'th
level with spin $\sigma$ for each wave vector $\mathbf{k}$. The
final states $\left|f_{\nu}\right\rangle $'s containing $\nu$ e-h
pairs are constructed by annihilating $\nu$ electrons in the occupied
conduction states and creating $\nu$ electrons in the unoccupied
conduction states from $\left|f_{0}\right\rangle $.

\subsection{Overlap integrals}

We assume that the transition matrix elements between the core state
and the photo-excited states are constant. The remaining matrix elements
connecting the ground and final states are expressed by \begin{equation}
\langle f_{\nu}|s_{\sigma}|g\rangle=\left|\begin{array}{cccc}
S_{1,1} & S_{2,1} & ..... & S_{N_{e},1}\\
S_{1,2} & S_{2,2} & ..... & S_{N_{e},2}\\
.... & .... & .... & ....\\
S_{1,N_{e}} & S_{2,N_{e}} & ...... & S_{N_{e},N_{e}}\end{array}\right|,\label{eq.overlap}\end{equation}
 with \begin{equation}
S_{i,i'}=\int\phi_{i}^{*}(\mathbf{r})\phi_{i'}^{(0)}(\mathbf{r}){\rm d}^{3}r,\end{equation}
 where the integral is carried out within a unit cell. Subscripts
$i=(\sigma,n,{\bf k})$ and $i'=(\sigma',n',{\bf k}')$ are running
over occupied conduction states in the presence of core hole and in
the ground state, respectively. We eliminate the overlaps between
the wave functions for the core levels. The corresponding energies
difference is given by

\begin{align}
\Delta E & =E_{f_{\nu}}-E_{g}\nonumber \\
 & =E_{f_{\nu}}-E_{f_{0}}+E_{f_{0}}-E_{g},\label{eq:ene_diff}\end{align}
where $E_{f_{0}}-E_{g}$ includes the energy of core hole, and is
treated as an adjustable parameter in the present study such that
the threshold of XPS spectra coincides with the experimental value.
The excitation energy $E_{f_{\nu}}-E_{f_{0}}$ with $\nu$ e-h pairs
is given by \begin{align}
E_{f_{\nu}}-E_{f_{0}} & =\sum_{\left(i,j\right)}(\epsilon_{j}-\epsilon_{i}),\label{eq:ene_eh}\end{align}
 where $\epsilon_{i}$'s are the Kohn-Sham eigenvalues, and $\epsilon_{j}-\epsilon_{i}$
stands for the energy of e-h pair of an electron at level $j$ and
a hole at level $i$. Although the Kohn-Sham eigenvalues may not be
proper quasi-particle energies, they practically give a good approximation
to quasi-particle energies, except for the fundamental energy gap.\cite{Hybertsen1986,Hamada1990}
Substituting Eqs.~(\ref{eq.overlap}) and (\ref{eq:ene_diff}) into
Eq.~(\ref{eq:xps}), we obtain the XPS spectra.

In the actual calculation, instead of $N_{c}$ ${\bf k}$-points,
we pick up only the $\Gamma$ point as the sample states for calculating
XPS spectra. For Ni, we pick up the $X$ point (and the equivalent
$Y$ and $Z$ points) in addition to the $\Gamma$ point, since the
$3d$ band states at the $\Gamma$ point are fully occupied by both
up-spin and down-spin electrons even though the $3\times3\times3$
fcc super-cell is used.

Before closing this section, we briefly mention the XPS intensity
at the energy of threshold. The final state $|f_{0}\rangle$ with
the lowest energy (no e-h pair) has a finite overlap with the ground
state $|g\rangle$, giving rise to intensities at the threshold. In
principle, such overlap converges at zero with $N_{e}\to\infty$,
according to the Anderson orthogonality theorem.\cite{Anderson1967}
In such infinite systems, energy levels become continuous near the
Fermi level and thereby infinite numbers of e-h pairs could be created
with infinitesimal excitation energies, leading to the so called Fermi
edge singularity in the XPS spectra. The finite contribution obtained
above arises from the discreteness of energy levels and could be interpreted
as the integrated intensity of singular spectra near the threshold,
in consistent with the model calculations for other systems.\citep{Kotani1974,Feldkamp1980}

\section{Results and Discussion}

\subsection{Ferromagnetic Transition Metals}

\begin{figure}[H]
 \hfill{}\includegraphics{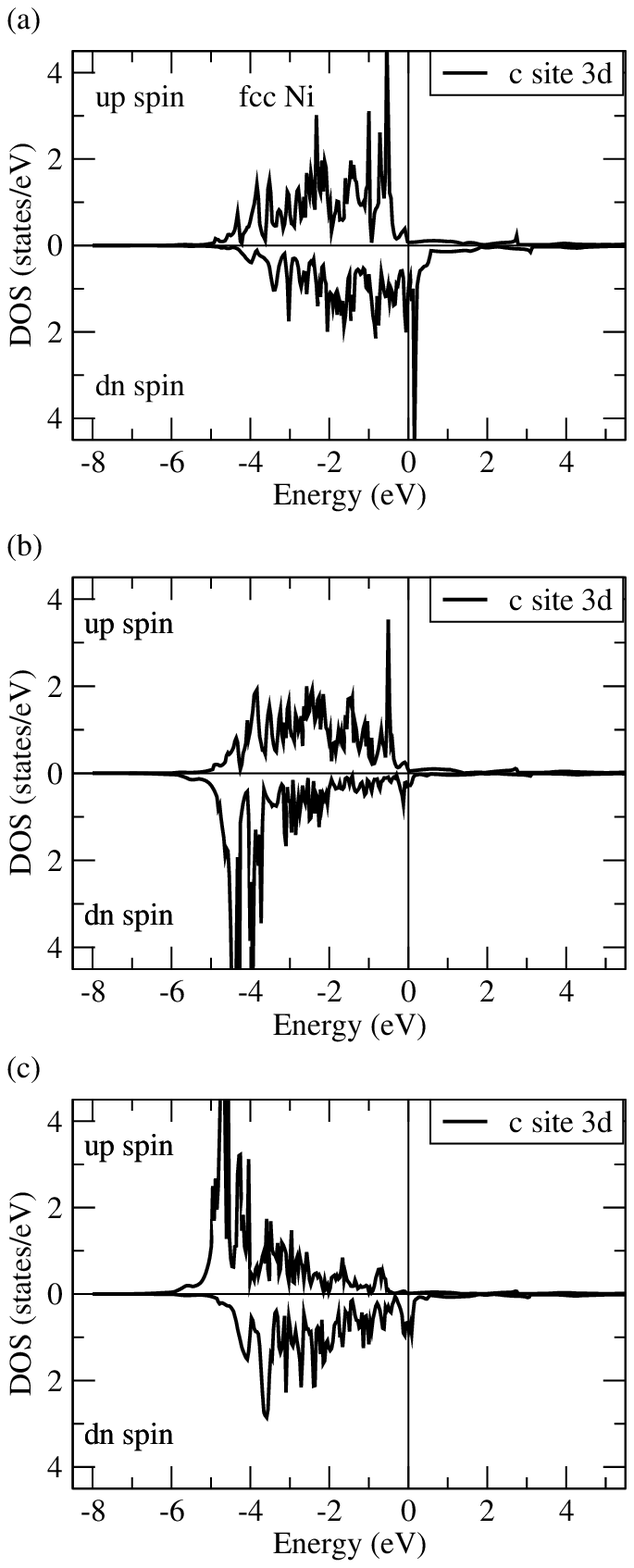}\hfill{}

\caption{Calculated $d$-DOS at the core-hole site in the super-cell system
in ferromagnetic nickel; (a) $d$-DOS with no core-hole, (b) $d$-DOS
when the $3s$ up-spin electron is removed, (c) $d$-DOS when the
$3s$ down-spin electron is removed. \label{fig:dos-fcc-ni}}

\end{figure}

\begin{figure}[H]
 \hfill{}\includegraphics{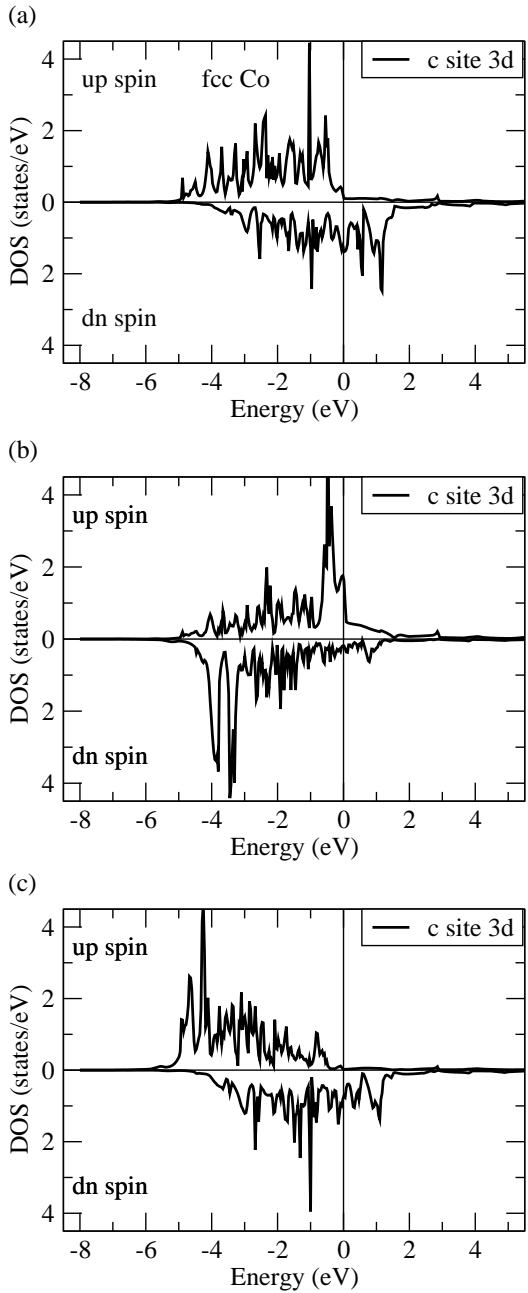}\hfill{}

\caption{Calculated $d$-DOS at the core-hole site in the super-cell system
in ferromagnetic cobalt; (a) $d$-DOS with no core-hole, (b) $d$-DOS
when the $3s$ up-spin electron is removed, (c) $d$-DOS when the
$3s$ down-spin electron is removed. \label{fig:dos-fcc-co}}

\end{figure}

In this subsections, we refer to majority(minority) spin as up(down)-spin.
The Ni metal takes an fcc structure. For simplicity, the Co metal
is assumed to take an fcc structure, although it actually takes an
hcp structure. The Fe metal takes a bcc structure. Figures \ref{fig:dos-fcc-ni}
and \ref{fig:dos-fcc-co} show DOS's projected onto the states with
$d$ symmetry ($d$-DOS) at the core-hole site for Ni and Co, respectively.
The corresponding DOS's for Fe are shown in Fig.~3 in Ref. \onlinecite{Taka2008FeXPS}.
In these calculations, six $k$-points are picked up in the irreducible
Brillouin zone for super-cell systems. The DOS's calculated with no
core-hole are essentially the same as those reported by Moruzzi, Janak
and Williams. \cite{Morruzi1978} Table \ref{table.1} lists the screening
electron number $\Delta n_{d\sigma}$ in the $d$-symmetric states
with spin $\sigma$, that is, the difference of the occupied electron
number between in the presence and in the absence of the core hole
inside the muffin-tin sphere.

On the basis of these one-electron states, we calculate the $3s$
XPS spectra, by following the procedure describe in Sec.~II. Figures
\ref{fig:XPS_Ni3s}, \ref{fig:XPS_Co3s} and \ref{fig:Fe3sXPS} are
the spectra thus calculated as a function of the binding energy $\omega_{q}-\epsilon$
for Ni, Co, and Fe, respectively, in comparison with the experiments.
\cite{Campen1994,See1995Ni,Xu1995} The spectral shape in Fig.~\ref{fig:Fe3sXPS}
for Fe is slightly different from our previous result (Fig.~4 in
Ref. \onlinecite{Taka2008FeXPS}), since the present calculation takes
full account of excitations up to three e-h pairs in comparison with
only up to two e-h pairs in Ref. \onlinecite{Taka2008FeXPS}. The
spectra are strikingly different between the up-spin channel and the
down-spin channel and strongly depend on elements, in good agreement
with the experiments. In the following, we explain the origin of these
behaviors in relation to one-electron states screening the core hole.

\begin{table}[tb]

\caption{Screening electron number in the $d$-symmetric states inside the
muffin-tin sphere at the $3s$ core-hole site. The radii of the muffin-tin
spheres are $2.0$ Bohr. }

\label{table.1}

\hfill{}\begin{tabular}{rrrrrrrrc}
\hline 
 &  & 3s hole spin  &  & $\Delta n_{d\uparrow}$  &  & $\Delta n_{d\downarrow}$  &  & $\Delta n_{d\uparrow}+\Delta n_{d\downarrow}$ \tabularnewline
\hline 
Fe  &  & up  &  & -1.44  &  & 2.38  &  & 0.94\tabularnewline
 &  & dn  &  & 0.47  &  & 0.48  &  & 0.95\tabularnewline
Co  &  & up  &  & -0.52  &  & 1.55  &  & 1.03\tabularnewline
 &  & dn  &  & 0.24  &  & 0.75  &  & 0.99\tabularnewline
Ni  &  & up  &  & 0.07  &  & 0.87  &  & 0.94\tabularnewline
 &  & dn  &  & 0.26  &  & 0.65  &  & 0.91\tabularnewline
\hline
\end{tabular}\hfill{}
\end{table}

\subsubsection{Up-spin channel}

First we consider the situation that a up-spin $3s$ electron is removed
in a unit cell. As shown in Figs. \ref{fig:dos-fcc-ni} (b) and \ref{fig:dos-fcc-co}(b),
the $d$-DOS at the core-hole site are strongly modified by the core-hole
potential for Ni and Co. The situation is similar to Fe, as shown
in Fig.~3 in Ref. \onlinecite{Taka2008FeXPS}. The $d$-DOS's of
the down-spin states are strongly pulled down, forming quasi-bound
states around the bottom of the $3d$ band. On the other hand, the
$d$-DOS's of the up-spin states are slightly pulled down for Ni,
and pushed upward to the higher energy region for Co and Fe. The $d$-DOS's
at the site without a core-hole are essentially the same as those
in the ground state with no core-hole

Since up-spin electrons are prevented from coming close to the core-hole
site due to the exchange interaction with the up-spin core hole, the
screening is almost done by down-spin electrons. This tendency is
clear in Ni; $\Delta n_{d\downarrow}=0.87$, while $\Delta n_{d\uparrow}=0.07$,
as shown in Table \ref{table.1}. For Co and Fe, the core-hole potential
is overscreened by down-spin electrons; $\Delta n_{d\downarrow}=1.55$(Co),
$\Delta n_{d\downarrow}=2.38$(Fe). This overscreening is compensated
by up-spin electrons; $\Delta n_{d\uparrow}=-0.52$(Co), $\Delta n_{d\uparrow}=-1.44$(Fe).
As a result, the screening electron numbers become almost unity: $0.94$,
$1.03$, and $0.94$ for Ni, Co, and Fe, respectively, indicating
that the screening is nearly completed by the $d$-symmetric states.
The magnetic moments at the core-hole site are $-0.2\,\mathrm{\mu_{B}}$,
$-0.5\,\mathrm{\mu_{B}}$, and $-1.8\,\mathrm{\mu_{B}}$ for Ni, Co,
and Fe, respectively, which are opposite to those without the core-hole,
$0.6\,\mathrm{\mu_{B}}$, $1.6\,\mathrm{\mu_{B}}$, $2.1\,\mathrm{\mu_{B}}$,
for Ni, Co, and Fe, respectively.

We note that the magnitude of the overscreening by the down-spin electrons
becomes weaker Fe, Co, and Ni in that order. In Ni, the number of
the down-spin electrons screening the core-hole is just about unity.
Consequently, the up-spin $3d$ states become to be less pushed upward
to the high energy region in that order. In Ni, the effects of the
core-hole potential and the Coulomb repulsion from the screening down
electrons subtly cancel out each other and the $d$-DOS of up-spin
states is hardly modified.

\begin{table}
\caption{Absolute squares $\left|A_{\uparrow}\right|^{2}$ and $\left|A_{\downarrow}\right|^{2}$,
where $A_{\uparrow}$ and $A_{\downarrow}$ represent the up- and
down-spin parts of the overlap integral, respectively, between $\left|f_{0}\right\rangle $
and $s_{\sigma}|g\rangle$ in the $\sigma$-spin channel, that is,
$\langle f_{0}|s_{\sigma}|g\rangle=A_{\uparrow}A_{\downarrow}$. \label{tab:ov-FeCoNi}}

\hfill{}\begin{tabular}{cccccccc}
\hline 
 &  & \multicolumn{2}{c}{up-spin core-hole} &  &  & \multicolumn{2}{c|}{dn-spin core-hole}\tabularnewline
\hline 
 &  & $\left|A_{\uparrow}\right|^{2}$  & $\left|A_{\downarrow}\right|^{2}$  &  &  & $\left|A_{\uparrow}\right|^{2}$  & $\left|A_{\downarrow}\right|^{2}$\tabularnewline
\hline 
Ni  &  & $0.940$  & $0.234$  &  &  & $0.770$  & $0.572$\tabularnewline
Co  &  & $0.983$  & $0.049$  &  &  & $0.980$  & $0.601$\tabularnewline
Fe  &  & $0.967$  & $0.127$  &  &  & $0.969$  & $0.909$\tabularnewline
\hline
\end{tabular}\hfill{}
\end{table}

It is inferred from these changes in the $d$-DOS's that one-electron
wave functions are largely modified by the core-hole potential particularly
for down-spin states. It is necessary to use both the occupied and
unoccupied states of the ground state in order to expand those modified
one-electron wave functions for the down-spin electrons, since the
down-spin $3d$ bands are partially occupied in the ground state.
For this reason, the absolute square $\left|A_{\downarrow}\right|^{2}$
becomes rather small, as shown in table \ref{tab:ov-FeCoNi}. Here
$A_{\uparrow}$ ($A_{\downarrow}$) represents the the up(down)-spin
part of the overlap integral between the lowest-energy final state
$|f_{0}\rangle$ (no e-h pair) and $s_{\uparrow}|g\rangle$, and thereby
$\langle f_{0}|s_{\uparrow}|g\rangle=A_{\uparrow}A_{\downarrow}$.

The one-electron wave functions for up-spin electrons are also modified
from those in the ground state. In spite of such modification, $\left|A_{\uparrow}\right|^{2}$'s
are nearly unity as shown in Table \ref{tab:ov-FeCoNi}. This could
be understood as follows. Since the up-spin $3d$ bands are almost
fully occupied in the ground state, up-spin one-electron states constituting
the final state $\left|f_{0}\right\rangle $ could be represented
by a unitary transform of those constituting the ground state $\left|g\right\rangle $.
Therefore, since the determinant is invariant under unitary transformation,
$A_{\uparrow}$'s are close to unity.

Final states $\left|f_{\nu}\right\rangle $'s containing \emph{up}-spin
e-h pairs could give rise to only small intensities, since the states
of the excited electrons with up-spin are almost orthogonal to occupied
states with up-spin in the ground state $\left|g\right\rangle $,
and thereby the overlap determinants would vanish. On the other hand,
the final states $\left|f_{1}\right\rangle $'s containing one \emph{down}-spin
e-h pair could give rise to considerable intensities, since the corresponding
one-electron wave functions contain the amplitudes of the unoccupied
one-electron states in the ground state $\left|g\right\rangle $,
and thereby the overlap determinants would not vanish. Considering
various combinations of one e-h pair, we obtain intensities distributed
in a wide range of binding energy.

\begin{figure}[H]
\hfill{}\includegraphics[scale=0.5]{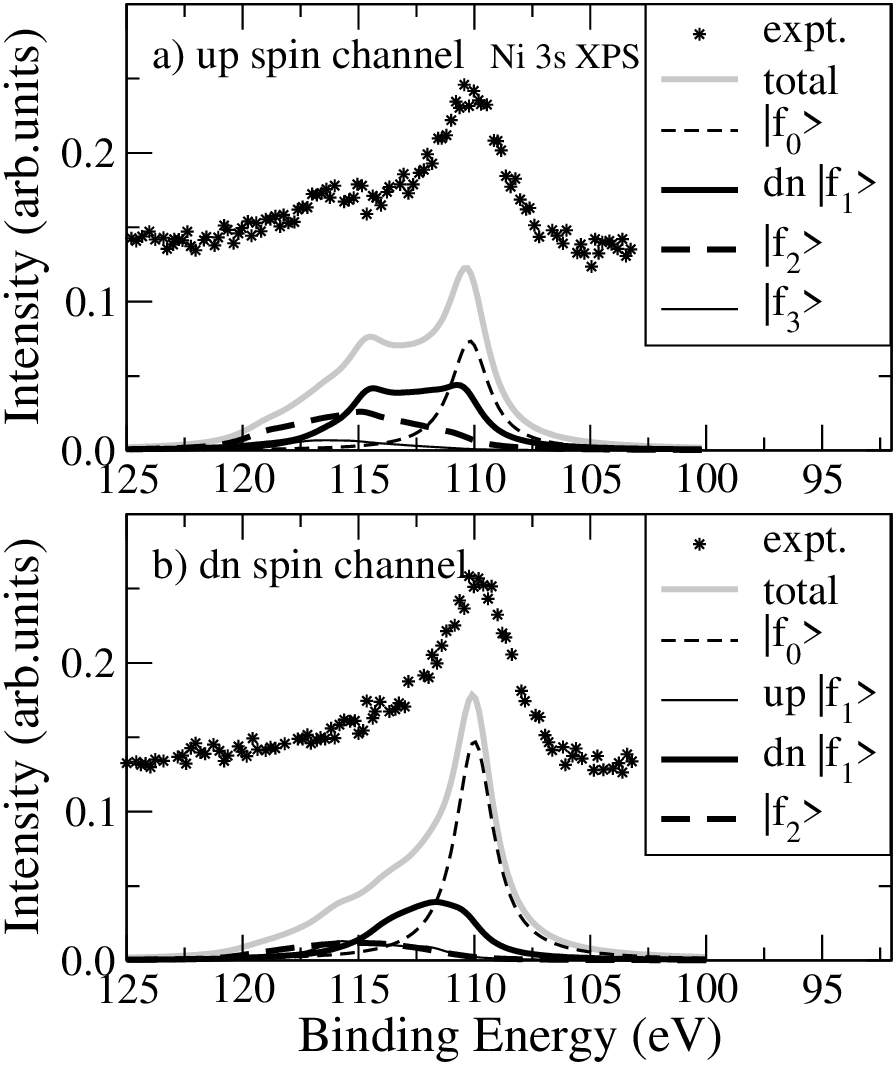}\hfill{}

\caption{$3s$ XPS spectra in ferromagnetic nickel as a function of binding
energy. (a) and (b) are for the up-spin and down-spin channels, respectively.
The experimental data are taken from Ref.~{[}\onlinecite{See1995Ni}{]}.\label{fig:XPS_Ni3s}}

\end{figure}

Figure \ref{fig:XPS_Ni3s}(a) shows the calculated spectra for Ni
in comparison with experimental observations in the up-spin channel.
The calculated spectra have maximum intensity at the threshold around
$\omega_{q}-\epsilon=110$ eV and the significant satellite intensity
around $\omega_{q}-\epsilon=115$ eV. Note that that the final states
containing one down-spin e-h pair give rise to considerable intensities,
extending over main and satellite regions. The satellite intensity
corresponds to excitations of one e-h pair from the quasi-bound states
to the unoccupied states with down-spin. This excitation may be considered
as a core-hole plus $d^{9}$, since one of the quasi bound states,
which are almost localized at the core-hole site, is empty. In this
calculation, however, the states are not split off from the band bottom
edge, indicating that the core-hole-dn-hole states are only weakly
bound. The satellite binding energy of the calculated spectra is $1\,\mathrm{eV}$
smaller than that of the observed spectra. This discrepancy might
be owing to the LDA. The Excitations of two e-h pairs would give rise
to considerable intensities in the energy range $\omega_{q}-\epsilon=110\sim120$
eV. Excitations of three e-h pairs give rise to only small intensities.

\begin{figure}[H]
\hfill{}\includegraphics[scale=0.5]{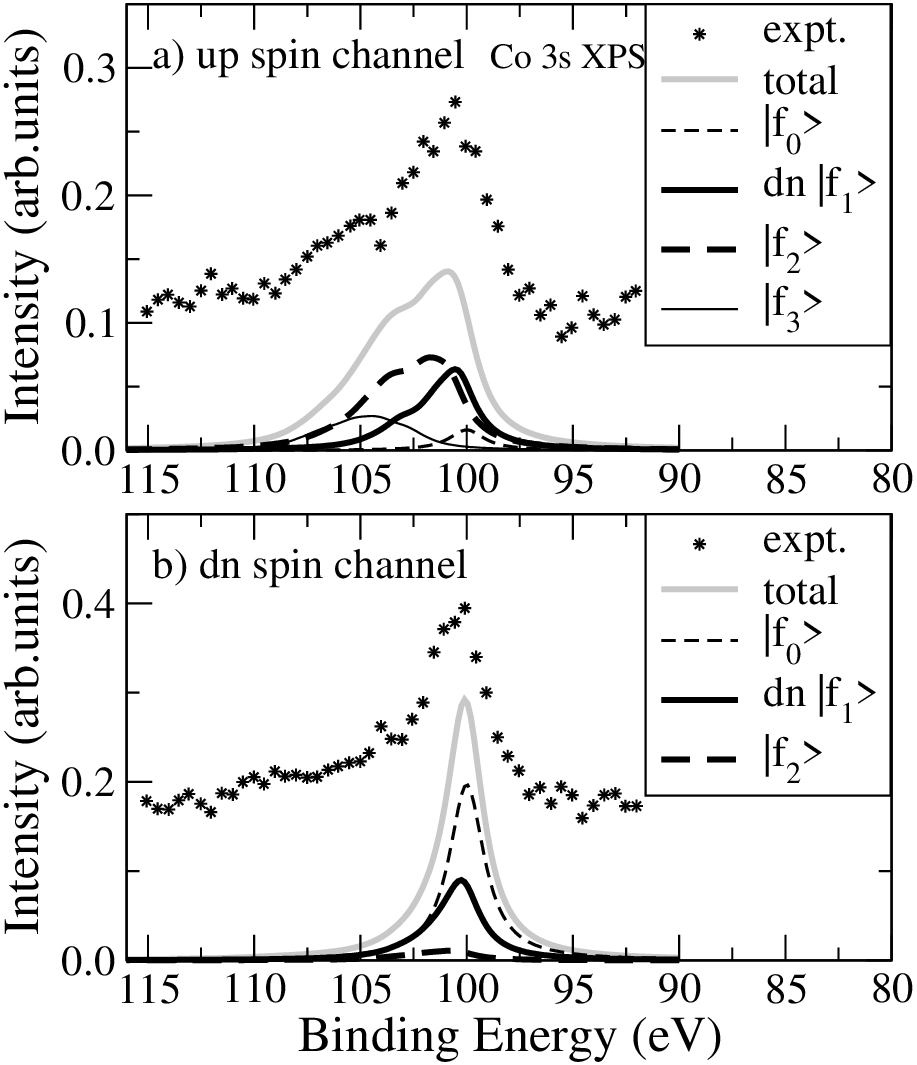}\hfill{}

\caption{$3s$ XPS spectra in ferromagnetic cobalt as a function of binding
energy. (a) and (b) are for the up-spin and down-spin channels, respectively.
The experimental data are taken from Ref.~{[}\onlinecite{Campen1994}\label{fig:XPS_Co3s}{]}.}

\end{figure}

Figure \ref{fig:XPS_Co3s}(a) shows the calculated spectra for Co.
The calculated spectra have a broad peak structure with the maximum
intensity at the threshold around $\omega_{q}-\epsilon=101$ eV. They
also have large shoulder intensities around $\omega_{q}-\epsilon=104$
eV. The former peak originates from the excitations with one and two
e-h pairs. The contribution of the lowest-energy final state $\left|f_{0}\right\rangle $
(no e-h pair) is quite small due to small $\left|A_{\downarrow}\right|^{2}$.
The latter shoulder originates from excitations of two and three e-h
pairs, probably including the excitations from the quasi-bound states
to the unoccupied states with down-spin.

\begin{figure}[H]
\hfill{}\includegraphics[scale=0.5]{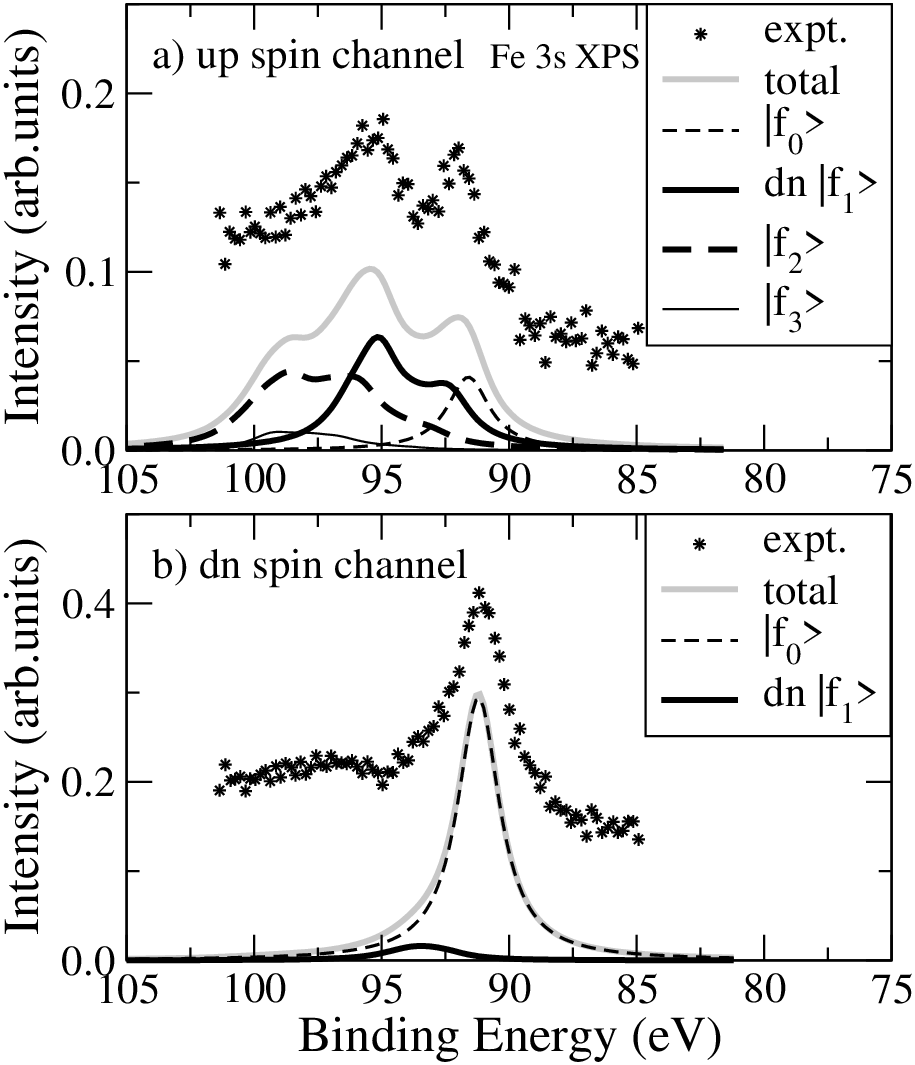}\hfill{}

\caption{$3s$ XPS spectra in ferromagnetic iron as a function of binding energy.
(a) and (b) are for the up-spin and down-spin channels, respectively.
The experimental data are taken from Ref.~{[}\onlinecite{Xu1995}\label{fig:Fe3sXPS}{]}. }

\end{figure}

Figure \ref{fig:Fe3sXPS}(a) shows the calculated spectra for Fe.
The calculated spectra consist of a peak around $\omega_{q}-\epsilon=92$
eV and a satellite peak around $95$ eV. The satellite peak is larger
than the peak around the threshold. The final states $\left|f_{1}\right\rangle $'s
containing one down-spin e-h pair give rise to the satellite intensity.
The final states $\left|f_{2}\right\rangle $'s containing two e-h
pairs give rise to a shoulder to the satellite around $\omega_{q}=\epsilon=97\sim100$
eV, as shown in the figure. The excitations of three e-h pairs gives
rise to finite but small intensities in the wide energy range around
$98$ eV.

\subsubsection{Down-spin channel}

When a down-spin $3s$ electron is removed in a unit cell, the screening
behavior is quite different from the situation where a up-spin $3s$
electron is removed. As shown in Figs. \ref{fig:dos-fcc-ni}(c) and
\ref{fig:dos-fcc-co}(c), and the bottom panel in Fig.~3 in Ref.
\onlinecite{Taka2008FeXPS}, the $d$-DOS's at the core-hole site
for Ni, Co, and Fe are strongly modified by the core-hole potential.
Different from the up-spin channel, the effect is larger for up-spin
conduction states than for down-spin conduction states; large weights
are transferred to the bottom of the conduction band in the up-spin
$d$-DOS's, while the weights are slightly shifted downward to the
lower energy region in the down-spin $d$-DOS's.

The screening electron numbers are rather smaller in the up-spin state
than in the down-spin state, as listed in Table \ref{table.1}. This
difference arises from the fact that up-spin $3d$ states are almost
occupied in the ground state. The total screening electron numbers
are almost unity, $0.91$, $0.99$, and $0.95$ for Ni, Co, and Fe,
respectively. The local magnetic moments at the core-hole site are
not significantly changed from those in the ground states.

The $|A_{\uparrow}|^{2}$'s are not far from unity as listed in Table
\ref{tab:ov-FeCoNi}, by the same reason as in the up-spin channel.
Note that $|A_{\downarrow}|^{2}$'s are larger than in the up-spin
channel, indicating that one-electron wave functions for down-spin
conduction electrons are less modified in the down-spin channel than
in the up spin-channel.

Figure \ref{fig:XPS_Ni3s}(b) shows the calculated spectra for Ni,
in comparison with the experiment. The final states $\left|f_{1}\right\rangle $'s
containing one e-h pair with down-spin give rise to considerable intensities
in a wide energy region $110\sim115$ eV, while the final states $\left|f_{2}\right\rangle $'s
containing two e-h pairs give rise to small intensities in a wide
energy region with the maximum around $\omega_{q}-\epsilon=115$ eV.
We obtain a asymmetric shape with a tail in the high-energy region
in agreement with the experiment.

Figure \ref{fig:XPS_Co3s}(b) shows the calculated spectra for Co.
The lowest-energy final states $\left|f_{0}\right\rangle $ gives
rise to a peak at the threshold around $\omega_{q}-\epsilon=100$
eV. Different from Ni, final states $\left|f_{1}\right\rangle $'s
and $\left|f_{2}\right\rangle $'s give rise to intensities on a limited
region near the threshold. This suggests that one-electron wave functions
are modified only for levels in the vicinity of the Fermi level. Contributions
of final states $\left|f_{3}\right\rangle $'s (three e-h pairs) are
found negligible.

Figure \ref{fig:Fe3sXPS}(b) shows the calculated spectra for Fe.
The final state $\left|f_{0}\right\rangle $ gives rise to a main
peak at the threshold $\omega_{q}-\epsilon=91$ eV with the largest
contribution in the three metals. Final states $\left|f_{1}\right\rangle $'s
with one down-spin e-h pair give rise to a broad peak around $94$
eV, which is quite small in comparison with the main peak. Although
the observed spectra show small satellite intensity around $\omega_{q}-\epsilon=97.5$
eV, the present calculation gives no intensity there. It is unclear
on the mechanism giving the intensity to our knowledge.

\subsection{Nonmagnetic metals}

In ferromagnetic metals Fe, Co, and Ni, the XPS spectral shapes remarkably
depend on the spin channel, owing to the exchange interaction between
the core hole and the conduction electrons. Although the spectral
shape is found closely related to the filling of band states, the
presence of the satellite in the XPS spectra seems to have no direct
relation to ferromagnetic states. In this subsection, taking up typical
nonmagnetic metals Cu, V and Ru, we clarify this issue. Since the
spectra are independent of the spin channel in nonmagnetic metals,
we consider the down-spin channel in the following. We assume the
fcc structure for Cu and Ru, and the bcc structure for V.

\begin{figure}[H]
\hfill{}\includegraphics[clip]{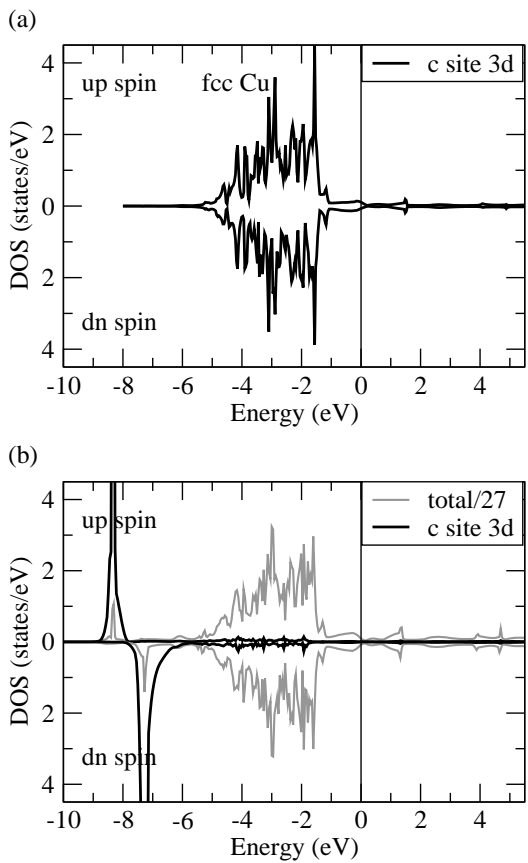}\hfill{}

\caption{(a) DOS calculated in a system of super-cell with no core-hole in
nonmagnetic copper. The solid line represents the DOS projected onto
the $d$-symmetry within the muffin-tin sphere. (b) DOS at the site
of $3s$ down-spin core-hole. The thin line represents the total DOS
divided by the number of atoms in a unit cell. \label{fig:DOS-Cu}}

\end{figure}

Figure \ref{fig:DOS-Cu} show the $d$-DOS for Cu. The DOS calculated
with no core-hole are essentially the same as those reported by Moruzzi,
Janak and Williams.\cite{Morruzi1978} The $d$-DOS at the core-hole
site is shifted downward to the deeper energy region. The localize
bound-states are clearly created below the $3d$ bands in both up-
and down-spin states. As listed in Table \ref{table.Screening-Cu-Mo},
the screening electron numbers are given by $\Delta n_{d\uparrow}=0.31$,
$\Delta n_{d\downarrow}=0.24$, and the total screening electron number
is given by $\Delta n_{d\uparrow}+\Delta n_{d\downarrow}=0.55$. This
indicates that the core-hole potential is not sufficiently screened
by the $3d$ electrons at the core-hole site. Here we note that the
screening mechanism seems to be somewhat different from the other
metals which have partially occupied $d$-bands and do not show the
bound states. Since the $3d$ band states are almost fully occupied
in the ground state, the change of the charge density is hardly achieved
by a unitary transform of the $3d$ band states. Since the bound states
are split off from the band bottom edge, the radial part of the local
atomic wavefunctions constituting the bound states at the core-hole
site is slightly shrunk compared to that in the ground state with
no core-hole, leading to the small change of the charge density. This
shrink of radial part of the local atomic wavefunctions cannot be
described by a unitary transform of the $3d$ band states in the ground
state. The states with much higher energy in the ground state necessarily
constitute the bound states to some extent.

\begin{figure}[H]
 \hfill{}\includegraphics[clip]{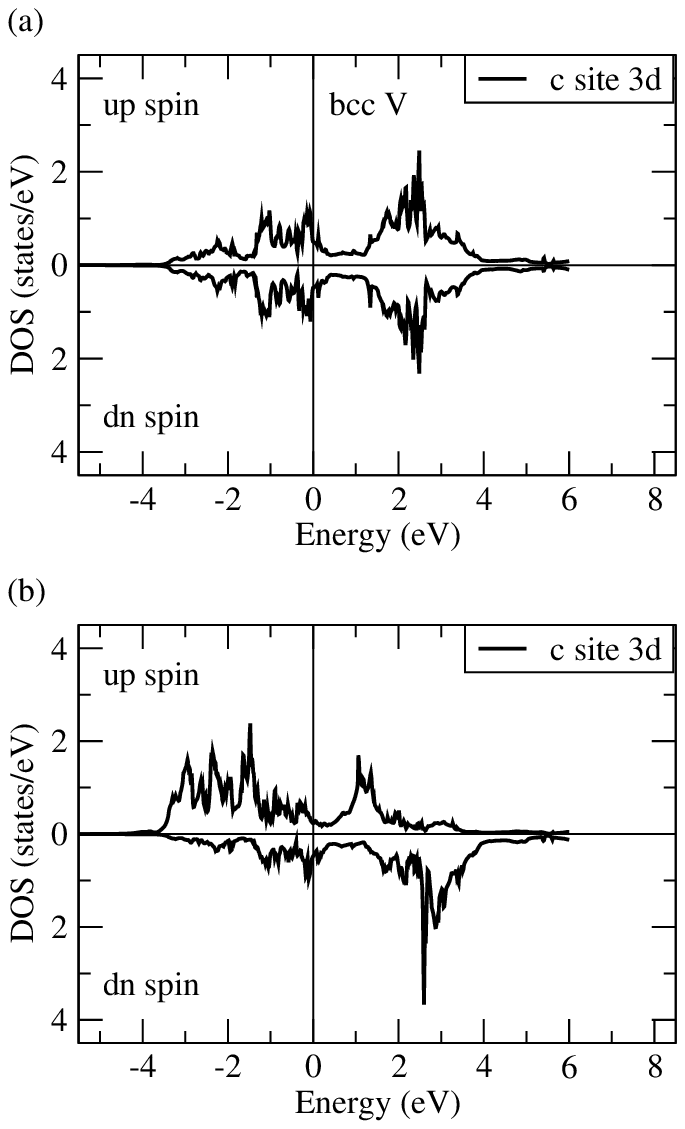}\hfill{}

\caption{(a) DOS calculated in a system of super-cell with no core-hole in
nonmagnetic vanadium. The solid line represents the DOS projected
onto the $d$-symmetry within the muffin-tin sphere. (b) DOS at the
site of $3s$ down-spin core-hole. \label{fig:DOS-V}}

The solid line represents the DOS projected onto the $d$-symmetry
within the muffin-tin sphere.
\end{figure}

Figure \ref{fig:DOS-V} shows the $d$-DOS for V. The $d$-DOS at
the core-hole site is shifted downward to the deeper energy region
for up-spin states, while the $d$-DOS is shifted upward for down-spin
states. No bound state is formed. As listed in Table \ref{table.Screening-Cu-Mo},
the screening electron numbers are given by $\Delta n_{d\uparrow}=1.21$,
$\Delta n_{d\downarrow}=-0.21$, and the total number by $\Delta n_{d\uparrow}+\Delta n_{d\downarrow}=1.00$.
The core-hole potential is overscreened by up-spin electrons, and
the overscreening is compensated by down-spin electrons. Note that
the screening is complete within the $3d$ electrons at the core-hole
site. The screening is effective because the $3d$ bands are partially
occupied.

\begin{figure}[H]
 \hfill{}\includegraphics[clip]{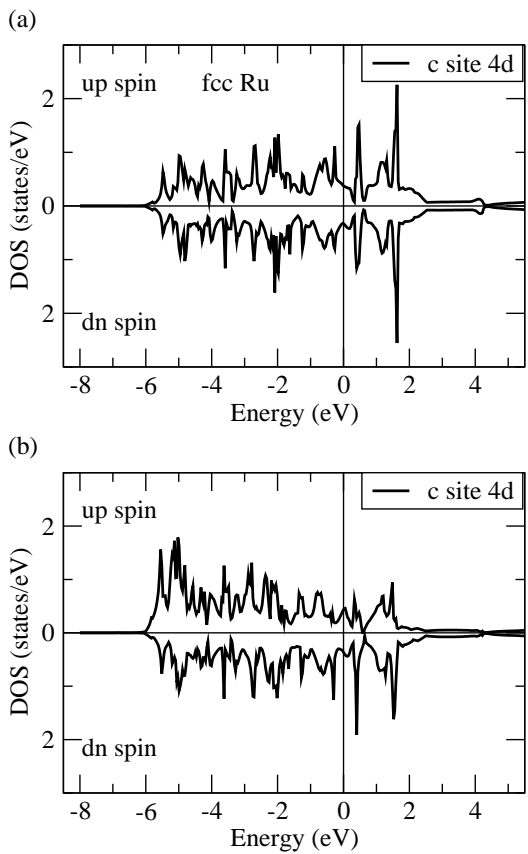}\hfill{}

\caption{(a) DOS calculated in a system of super-cell with no core-hole in
nonmagnetic ruthenium. The solid line represents the DOS projected
onto the $d$-symmetry within the muffin-tin sphere. (b) DOS at the
site of $4s$ down-spin core-hole. \label{fig:DOS-Ru}}

\end{figure}

Figure \ref{fig:DOS-Ru} shows the $d$-DOS for Ru. The $d$-DOS's
at the core-hole site are shifted downward to the deeper energy region
with both spin states, but the change is the smallest in the three
cases without any bound states. This is probably related to the fact
that the $4d$ electrons are more itinerant than the $3d$ electrons.
As listed in Table \ref{table.Screening-Cu-Mo}, the screening electron
numbers are $\Delta n_{d\uparrow}=0.77$, $\Delta n_{d\downarrow}=0.24$,
and $\Delta n_{d\uparrow}+\Delta n_{d\downarrow}=1.01$, indicating
that the screening is nearly completed by the $4d$ electrons at the
core-hole site.

%
\begin{table}[H]

\caption{Screening electron number with respect to the $d$ symmetry within
the muffin-tin sphere at the $3s$ ($4s$) down-spin core-hole site.
The radii of the muffin-tin spheres are $2.0$, $2.1$, and $2.3$
Bohr for Cu, V, and Ru, respectively.\label{table.Screening-Cu-Mo}}

\hfill{}\begin{tabular}{rrrrrrrrc}
\hline 
 &  & 3s hole spin &  & $\Delta n_{d\uparrow}$  &  & $\Delta n_{d\downarrow}$  &  & $\Delta n_{d\uparrow}+\Delta n_{d\downarrow}$ \tabularnewline
\hline 
Cu  &  & dn  &  & $0.31$  &  & $0.24$  &  & $0.55$\tabularnewline
V  &  & dn  &  & $1.44$  &  & $-0.44$  &  & $1.00$\tabularnewline
Ru  &  & dn  &  & $0.77$  &  & $0.24$  &  & $1.01$\tabularnewline
\hline
\end{tabular}\hfill{}
\end{table}

We calculate the up- and down-spin parts of overlap integrals between
the lowest-energy final state $\left|f_{0}\right\rangle $ and the
ground state $|g\rangle$, which values are listed in Table \ref{tab:ov-CuVRu}.

\begin{table}[H]
\caption{Absolute squares $\left|A_{\uparrow}\right|^{2}$ and $\left|A_{\downarrow}\right|^{2}$,
where $A_{\uparrow}$ and $A_{\downarrow}$ represent the up- and
down-spin parts of the overlap integral between $\left|f_{0}\right\rangle $
and $s_{\downarrow}|g\rangle$, that is, $\langle f_{0}|s_{\downarrow}|g\rangle=A_{\uparrow}A_{\downarrow}$.
\label{tab:ov-CuVRu}}

\hfill{}\begin{tabular}{ccccc}
\hline 
 &  & \multicolumn{3}{c}{dn-spin core-hole}\tabularnewline
\hline 
 &  & $\left|A_{\uparrow}\right|^{2}$  &  & $\left|A_{\downarrow}\right|^{2}$ \tabularnewline
\hline 
Cu  &  & $0.910$  &  & $0.935$ \tabularnewline
V  &  & $0.571$  &  & $0.938$ \tabularnewline
Ru  &  & $0.681$  &  & $0.949$ \tabularnewline
\hline
\end{tabular}\hfill{}
\end{table}

\begin{figure}[H]
\hfill{}\includegraphics[scale=0.3]{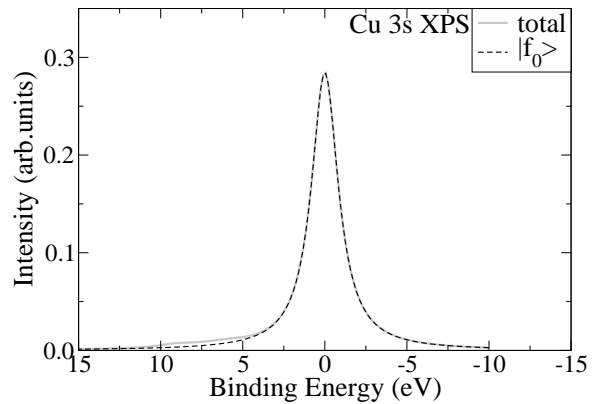}\hfill{}

\caption{$3s$ XPS spectra in nonmagnetic copper as a function of binding energy
in the down-spin channel.\label{fig:XPS-Cu}}

\end{figure}

For Cu, $|A_{\uparrow}|^{2}$ and $|A_{\downarrow}|^{2}$ are close
to unity, although the one-electron wave functions are strongly modified.
Since the $3d$ bands are fully occupied in the ground state, one-electron
wave functions constituting $\left|f_{0}\right\rangle $ are nearly
expressed by a unitary transform of those constituting $|g\rangle$,
except for the shrink of the radial part of the atomic wavefunctions
at the core-hole site. Therefore, the squares of the overlap determinant
$|A_{\uparrow}|^{2}$ and $|A_{\downarrow}|^{2}$ are nearly unity.
Final states $\left|f_{1}\right\rangle $ (one e-h pair), $\left|f_{2}\right\rangle $
(two e-h pairs), and so on, could have merely very small overlaps
with $s_{\downarrow}|g\rangle$, since the one-electron state on which
the excited electron sits is nearly orthogonal to the one-electron
states constituting the ground state. Thus we have a simple single
peak structure coming from $\left|f_{0}\right\rangle $ without any
noticeable intensities on the higher binding energy side, as shown
in Fig.~\ref{fig:XPS-Cu}. This result is consistent with the experimental
observation.\cite{Fadley1970} Note that the change of the wave functions
is not directly related to the spectral shape. The effect due to forming
strong bound states is not seen.

\begin{figure}[H]
\hfill{}\includegraphics[scale=0.3]{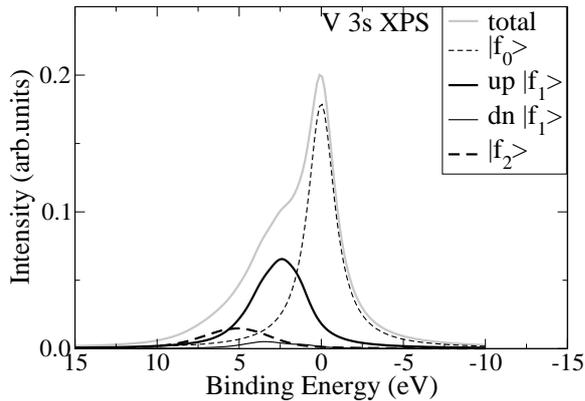}\hfill{}

\caption{$3s$ XPS spectra in nonmagnetic vanadium as a function of binding
energy in the down-spin channel.\label{fig:XPS-V}}

\end{figure}

For V, $\left|A_{\downarrow}\right|^{2}$ is close to unity. This
suggests that one-electron wave functions for down-spin electrons
are little modified from those in the ground state. On the other hand,
$\left|A_{\uparrow}\right|^{2}$ is rather smaller than unity. This
suggests that up-spin one-electron wave functions constituting $\left|f_{0}\right\rangle $
include the amplitudes of the unoccupied one-electron states in the
ground state. In such a situation, the final states $\left|f_{1}\right\rangle $'s
containing one e-h pair with up-spin could have finite overlaps with
the ground state. Figure \ref{fig:XPS-V} shows the calculated spectra.
We have a main peak coming from the final state $\left|f_{0}\right\rangle $
at the threshold and the noticeable shoulder coming from the final
states $\left|f_{1}\right\rangle $'s.

For Ru, $\left|A_{\downarrow}\right|^{2}$ is again close to unity
by the same reason as in V. The $\left|A_{\uparrow}\right|^{2}$ is
smaller than unity, although it is larger than that in V. This indicates
that one-electron wave functions constituting $\left|f_{0}\right\rangle $
are less modified by the core-hole potential in comparison with V.
Final states $\left|f_{1}\right\rangle $'s containing one e-h pair
with up-spin give rise to a shoulder structure with a little smaller
intensity than in V.

\begin{figure}[H]
\hfill{}\includegraphics[scale=0.3]{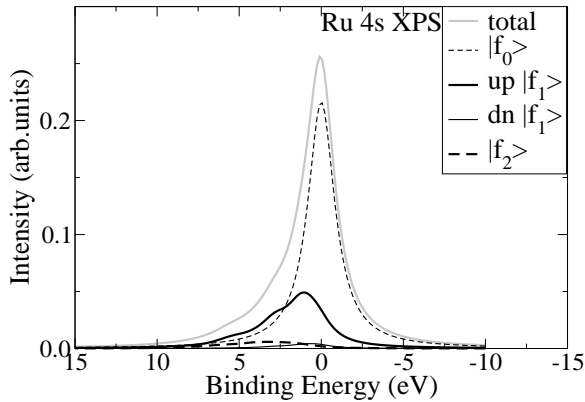}\hfill{}

\caption{$4s$ XPS spectra in nonmagnetic ruthenium as a function of binding
energy for the $4s$ down-spin core-hole.\label{fig:XPS-Ru}}

\end{figure}

\section{Concluding Remarks}

We have developed an \emph{ab initio} method to calculate the $3s$
and $4s$ core-level XPS spectra in ferromagnetic metals Ni, Co, and
Fe, and in nonmagnetic metals Cu, V, and Ru. For the ferromagnetic
metals, we have found that the spectral intensities are distributed
in a wide range of binding energy with satellite or shoulder structures
for the up-spin channel, while the intensities are concentrated near
the threshold with no satellite peak for the down-spin channel. The
origin of such behavior has been explained in relation to the $3d$
band modified by the core-hole potential and the overlap integral
between the final states and the ground state. Bound or quasi-bound
states are formed by the core-hole potential, and the e-h excitations
from such quasi-bound states to the unoccupied levels would usually
give rise to satellite intensities. However, the presence of the quasi-bound
state is not a sufficient condition to the presence of satellite;
the $d$-band should be partially occupied in the ground state, and
thereby the one-electron wave functions constituting the final states
include the amplitudes of the unoccupied one-electron states in the
ground state. If the $d$-band is fully occupied, the satellite intensity
would not come out even in the presence of the bound state. Note that
the satellite peak position has no direct relation to the $3s$ level
exchange splitting; the LDA calculation gives such splittings as $0.7$,
$1.9$, and $2.5$ eV for Ni, Co, and Fe, respectively.

These results indicate that the presence of satellite is not directly
related to the ferromagnetic ground state. We have clarified this
point by calculating the spectra in nonmagnetic metals Cu, V, and
Ru. For V and Ru, we have obtained shoulder structures in the XPS
spectra, although the structure is rather small for Ru. The origin
of these behaviors is the same as in the ferromagnetic metals. For
Cu, only a symmetric peak is found with no structure, although the
\emph{localized} bound states are clearly formed below the bottom
of the conduction band. This is because the $3d$ band is completely
occupied in the ground state.

We have calculated the XPS spectra in Ni, Co, Fe, and Cu in good agreement
with the experiment, while we could not find experimental XPS data
for V and Ru. Acker et al. observed the satellite structures even
in some Pauli paramagnetic Fe compounds,\cite{Acker1988} The present
results would provide an interpretation of their findings. Finally,
as regards the $L$-edge spectra, experimental data for XPS spectra
and the x-ray absorption spectra are accumulated, and \emph{ab initio}
approach has been tried.\cite{Kruger2004} The extension of the present
method to the $L$-edge spectra is left in future study.
\begin{acknowledgments}
We used the FLAPW code developed by Noriaki Hamada. We thank him for
allowing us to use his code and fruitful discussions. This work was
partially supported by a Grant-in-Aid for Scientific Research in Priority
Areas \textquotedblleft{}Development of New Quantum Simulators and
Quantum Design\textquotedblright{} (No.19019001) of The Ministry of
Education, Culture, Sports, Science, and Technology, Japan.
\end{acknowledgments}
\bibliographystyle{apsrev} \bibliographystyle{apsrev}
\bibliography{Bibcore}

\end{document}